\documentclass[oldversion, longauth, letter]{aa}  

\usepackage{scrextend}
\usepackage[colorlinks]{hyperref}

\usepackage{booktabs}
\usepackage{tabularx} 
\let\orgautoref\autoref
\renewcommand{\autoref}
        {\def\equationautorefname{Eq.}%
         \def\figureautorefname{Fig.}%
         \def\sectionautorefname{Sect.}%
         \def\subsectionautorefname{Sect.}%
         \def\subsubsectionautorefname{Sect.}%
         \orgautoref}

\usepackage{xcolor}
\definecolor{dark-red}{rgb}{0.9,0.0,0.0}
\definecolor{dark-blue}{rgb}{0.15,0.15,0.9}
\definecolor{dark-green}{rgb}{0.15,0.8,0.15}
\definecolor{medium-blue}{rgb}{0,0,0.9}
\definecolor{pink}{rgb}{1.0, 0.255, 0.831}
\hypersetup{linkcolor={dark-blue},citecolor={dark-blue}, urlcolor={medium-blue}
}

\makeatletter
\renewcommand*\aa@pageof{, page \thepage{} of \pageref*{LastPage}} 
\makeatother

\usepackage{changepage} 
\usepackage{adjustbox}
\usepackage{rotating}

\usepackage{graphicx}
\usepackage{txfonts}
\usepackage{soul}
\usepackage{url}
\usepackage{siunitx}
\usepackage{makecell}
\usepackage[nameinlink,capitalize]{cleveref}
\crefname{section}{Sect.}{Sect.}
\Crefname{section}{Section}{Section}

\begin{document}

    \title{$^{13}$CO and potential variability in $\beta$ Pictoris b with GRAVITY+}

%

    \author{A.\,von Stauffenberg\inst{1} 
    \and J.\,Sauter\inst{1}
    \and P.\,Molli\`ere\inst{1} 
    \and M.\,Ravet\inst{1,3,4}
    \and D.\,Trevascus\inst{1}
    \and W.\,Brandner\inst{1}
    \and A.\,Berdeu\inst{5}
    \and M.\,Bonnefoy\inst{4}
    \and G.\,Bourdarot\inst{6}
    \and J.-B\,Le Bouquin\inst{2}
    \and G.\,Chauvin\inst{1,3}
    \and F.\,Eisenhauer\inst{6}
    \and M.\,Houllé\inst{2}
    \and L.\,Kreidberg\inst{1}
    \and E.\,Matthews\inst{1}
    \and F.\,Millour\inst{3}
    \and J.\,Scigliuto\inst{3}
    \and J.\,Wang\inst{7}
    \and J.W.\,Xuan\inst{8,9}
    \and Y.\,Zhang\inst{8}
    \and GRAVITY+ Collaboration}

    \institute{Max-Planck-Institut f\"ur Astronomie, K\"onigstuhl 17, 69117 Heidelberg, Germany 
    \and Université Grenoble Alpes: Saint-Martin-d'Hères, Auvergne-Rhône-Alpes, France
    \and Laboratoire J.-L. Lagrange, Université Côte d’Azur, Observatoire de la Côte d’Azur, CNRS, 06304 Nice, France
    \and Institut de Planétologie et d’Astrophysique de Grenoble (IPAG), Université Grenoble Alpes, CS 40700,38058 Grenoble Cédex 9, France
    \and LESIA, Observatoire de Paris, Université PSL, Sorbonne Université, Université Paris Cité, CNRS, 5 place Jules Janssen, 92195 Meudon, France
    \and Max-Planck-Institut f\"ur extraterrestrische Physik, Gießenbachstraße 1, 85748 Garching bei M\"unchen, Germany
    \and Center for Interdisciplinary Exploration and Research in Astrophysics (CIERA) and Department of Physics and Astronomy, Northwestern University, Evanston, IL 60208, USA
    \and Department of Astronomy, California Institute of Technology, Pasadena, CA 91125, USA
    \and Department of Earth, Planetary, and Space Sciences, University of California, Los Angeles, CA 90095, USA}

   \date{Received XXX / Accepted XXX}

 
  \abstract 
   {The $^{12}$CO/$^{13}$CO ratio was introduced as an indicator for where in the disk a planet has formed. Previously a lower value compared to the host star's was suggested to show that a planet accreted CO ice beyond the disk’s CO ice line. In this letter we aim to determine the $^{12}$CO/$^{13}$CO value of the directly imaged planet $\beta$ Pictoris b, and whether we can link it to its formation. Its apparent brightness results in an exceptional S/N of up to $\sim$60 per wavelength point. We present the first science observations with the upgraded GRAVITY+ instrument at a spectral resolution of R $\approx$ 4000, which we analyse with {\tt petitRADTRANS}. Our retrievals robustly indicate the presence of $^{13}$CO with a $^{12}$CO/$^{13}$CO ratio of 91$^{+24}_{-17}$, consistent with both a solar to ISM-like value. Our $^{12}$CO/$^{13}$CO value corroborates recent interpretations that $^{13}$CO may be a less useful tracer of formation location in the disk than previously thought; nonetheless, we discuss theories with which this value is consistent. As our observations span $\approx$ 7 hours, this enabled us to search for atmospheric variability in $\beta$ Pictoris b; we report a tentative constraint on the variability amplitude of about 1.4$^{+0.6}_{-0.7}$\%.}


\authorrunning{von Stauffenberg et al.}
\titlerunning{Analysis of $\beta$ Pic b}

\maketitle


\section{Introduction}
\label{sec:Introduction}

Isotopologues are molecules that only differ in the isotopes of their constituent atoms. A commonly studied example are the carbon monoxide isotopologues, $^{12}\text{C}^{16}\text{O}$ and $^{13}\text{C}^{16}\text{O}$. Here we focus on $^{13}\text{C}^{16}\text{O}$, which, for substellar atmospheres, has been discussed in many different works \citep[e.g.,][]{molliere2019a,zhang2021a,zhang2021b, deregt2024, gonzalespicos2024, zhang2024, gonzalesicos2025a, ghandi2025, deregt2025, grasser2025, deregt2026, ruffio2026, xuan2026}. $^{13}\text{CO}$ has been suggested to trace the formation history of giant planets. For example, a planet with an enriched $^{13}\text{CO}$ content was interpreted to have accreted significant CO ice, since CO ice was conjectured to be $^{13}\text{C}$-rich \citep{zhang2021a}. Since, beyond the CO snowline, the bulk of the carbon reservoir is frozen out \citep{oeberg2011, molliere2022}, such a fractionation may be unlikely. However, since \cite{zhang2021a}, most reported $^{12}\text{CO}/^{13}\text{CO}$ values for substellar objects have been consistent with ISM and solar values, including YSES-1\,b's updated constraint \citep{zhang2024}, and the ESO SupJup survey (refs.\ above). Here we investigate the $^{12}\text{CO}/^{13}\text{CO}$ ratio of the directly imaged planet $\beta$ Pic b. \cite{ravet2025} reported the first tentative measurement of $\log^{12}\text{CO}/^{13}\text{CO}$ ($1.12^{+0.11}_{-0.08}$), but caution that the value is likely affected by tellurics. In this letter, we revisit $\beta$ Pic b with the now updated version of the AO system of GRAVITY \citep{gravity_collab2026}, in order to better constrain the $^{12}\text{CO}/^{13}\text{CO}$ ratio.

\begin{figure*}[!h]
\centering
\includegraphics[width=\linewidth]{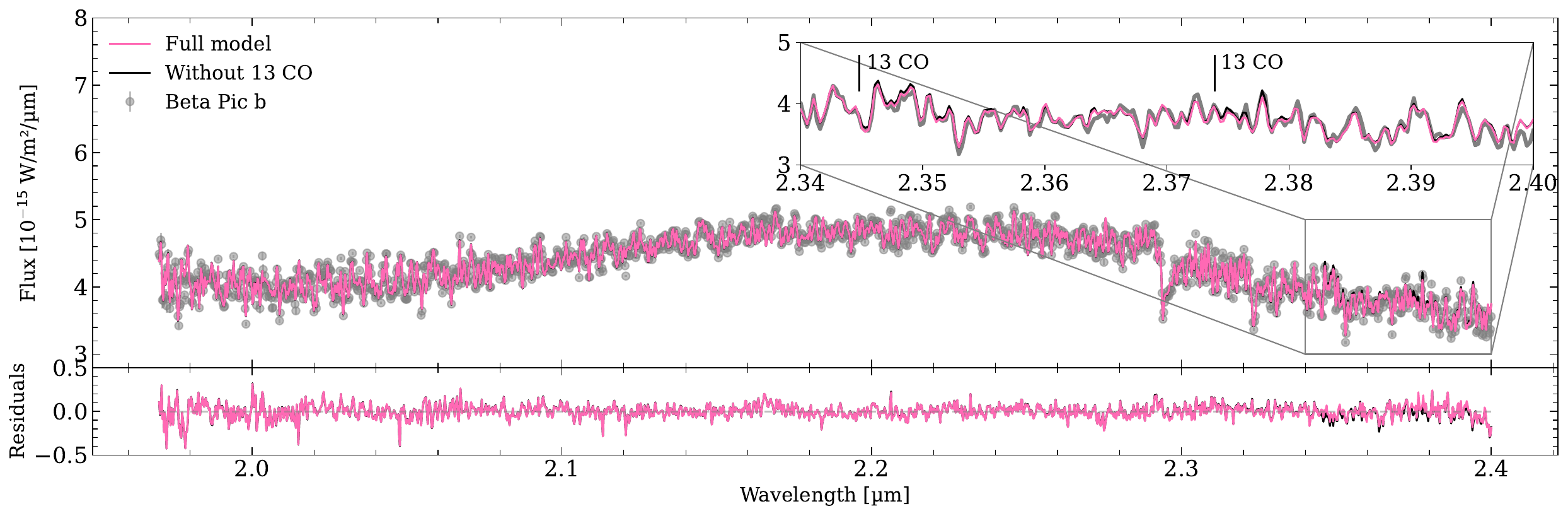}
\caption{{\em Top panel}: Shows the median K-band spectrum for GRAVITY+ (grey) with a median SNR of $\approx 180$. Its best-fit retrieval, including $^{13}$CO, is shown in pink, and the retrieval without $^{13}$CO in black. The inset shows a zoomed-in version of the 2.34 to 2.40 µm region. {\em Bottom panel}: The residuals between data and each model, respectively. The retrieval inflates the error bars in order to find the best fit model, resulting in residuals not explained by the model.}
\label{fig: best_fit}
\end{figure*}

Our secondary goal is to constrain possible spectroscopic variability in $\beta$ Pic b. Brown dwarfs are known to have significant variability driven by structural changes in their upper atmospheres. This may be caused by the sinking of silicate clouds, causing an inhomogeneous cloud cover, thought to govern the L/T transition \citep{radigan2014}, or chemical instabilities \citep{mccarthy2025}. To study variability in a bona-fide planet, $\beta$ Pic b is an ideal candidate, due to its K-band brightness of 12.43$\pm$0.07 \citep{males2014}. However, it is an analogue for early to mid L-dwarfs, which on average show $\leq$3\% variability \citep{crossfield2014}. This low predicted amplitude is compensated by its orbital inclination $89\pm0.01^{\circ}$ and short expected rotation period of $\sim$8.7\,hrs \citep{nowak2020, landman2024}. Assuming spin-orbit alignment, this could indicate an equator-on viewing geometry that would maximize rotational flux modulations.

\section{Observations and data reduction}
\label{sec:Observations}

As part of the GRAVITY+ GTO program 114.27JS (PI: Kreidberg), we obtained 7\,hrs of observations of the $\beta$ Pictoris system on 2024-12-20 using all four 8\,m Unit Telescopes in high-resolution mode (R$_{\lambda}\approx 4000$) with the dual-field on-axis configuration. The data were reduced with the ESO GRAVITY pipeline (v1.9.4) to obtain complex visibilities, and further processed with the exogravity\footnote{\url{https://gitlab.obspm.fr/mnowak/exogravity}}
pipeline to extract planet contrast spectra \citep{gravity_collab2019}. The stellar signal was removed using PHOENIX NewEra models \citep{hauschildt1997, hauschildt1999, hauschildt2025}, adopting T$_{\text{eff}}$ = 8000\,K, $\log\,\text{g}$ = 4.00, $[\text{M/H}]$ = 0.00, and $\alpha$/H = 0.2 \citep{swastik2021, reggiani2024}. Rotational broadening of 130\,km\,s$^{-1}$ \citep{royer2004} was applied using \cite{carvalho2023}, along with a Doppler shift of $-35$\,km\,s$^{-1}$ derived from $\chi^2$ minimisation of the Brackett-$\gamma$ feature in the corrected contrast spectra and stellar model. As these are ground-based observations, careful treatment of the tellurics is required. We apply a novel correction based on Beer–Lambert’s law, linearly interpolating flux following the airmass over the night, as described in ~\ref{app: telluric_correction}. To improve S/N, we combine our epochs (see \autoref{app: mean_averaging}), with the final spectrum shown in Fig.~\ref{fig: best_fit}.

\section{Methods}
\label{sec:Methods}

\subsection{Atmospheric modelling}

In order to characterise $\beta$ Pic b's atmosphere, we conducted retrievals using {\tt petitRADTRANS} \citep{molliere2019b, blain2024, nasedkin2024a}. In retrievals, a forward model for the spectra is repeatedly evaluated to identify the combinations of atmospheric parameters that reproduce the observed spectrum. Rather than in a self-consistent modelling approach, as described in \autoref{app: exorem-method}, retrievals treat various processes through free parameters and infer their posterior distributions given the data \citep[e.g.,][]{madhusudhan2018}. We use a combination of equilibrium and free chemistry. Most chemical species were interpolated using equilibrium tables, while major radiation absorbing species were freely retrieved. We list the priors of all parameters in \autoref{tab: priors_posteriors}. The detailed forward model is described in \autoref{app:forward model}. Additionally, we cross-correlate template spectra over the $^{13}$CO absorption region from 2.34 to 2.40\,µm. In order to construct the $^{13}$CO template, we follow \cite{molliere2019a,zhang2021a} and subtract the model neglecting $^{13}$CO from the best-fit retrieval model. The same subtraction is applied to the data to obtain $^{13}$CO residuals, while full-model residuals are computed by subtracting the complete model. Prior to cross-correlation, templates are high-pass filtered with a Gaussian kernel \citep{zhang2021a}, with the kernel width optimised to maximise the signal-to-noise. This balances noise suppression against over-smoothing of the signal. The residuals are weighted by their measurement uncertainties, ensuring appropriate statistical significance across wavelengths and minimising the contribution of poorly constrained flux points in the CCF signal.

\subsection{Variability analysis}
\label{sec: variability-analysis}

A common method to search for variability in time-domain data is via spectral light curves \citep[e.g.,][]{biller2013}. Each spectrum was first cleaned using iterative 3$\sigma$ clipping to remove instrumental outliers (e.g., bad pixels and residual noise). Our data are affected by strong chromatic variations from fibre coupling (see \autoref{fig:lomb_scargle} and \citealt{sauter2026}); these were removed by dividing each spectrum by the time-averaged mean and fitting a second-order polynomial. Corrected spectra were obtained by dividing the original spectra by this fit, acting as a high-pass filter to remove low-order continuum variations, using a 1D Gaussian kernel of standard deviation $\sigma = 3$ pixels along the wavelength axis. As we no longer have access to continuum variability, we analysed variability in the most prominent spectral features, focusing on the $^{12}$CO bandheads ($\sim 2.29$, 2.32, 2.35\,µm), which can be affected by chemical instabilities or clouds  \citep{oliveros_gomez2026}. We restricted our analysis to 2.05–2.355\,µm, which still covers all three bandheads. Below 2.05\,µm, both telluric CO$_2$ and H$_2$O begin to absorb, and beyond 2.35\,µm, CH$_4$ and H$_2$O absorb. Additionally, at long wavelengths, thermal background adds noise. We applied generalized Lomb-Scargle (GLS) periodograms \citep{zechmeister2009} to search for periodicities near the proposed rotational period of $\approx 8.7$\,hrs and its harmonics. The bandhead analysis regions span 2.281–2.296\,µm, 2.317–2.325\,µm, and 2.347–2.352\,µm, covering each absorption line and its onset. We then applied sinusoidal fits to the CO bandhead light curves, with the expected rotation period as an initial guess, to obtain the median periods and amplitudes.

\section{Results}
\label{sec:Results}

\subsection{Atmospheric analysis}
\label{sec: retrieval_results}

\begin{figure}[t!]
    \centering
    \includegraphics[width=0.98\linewidth]{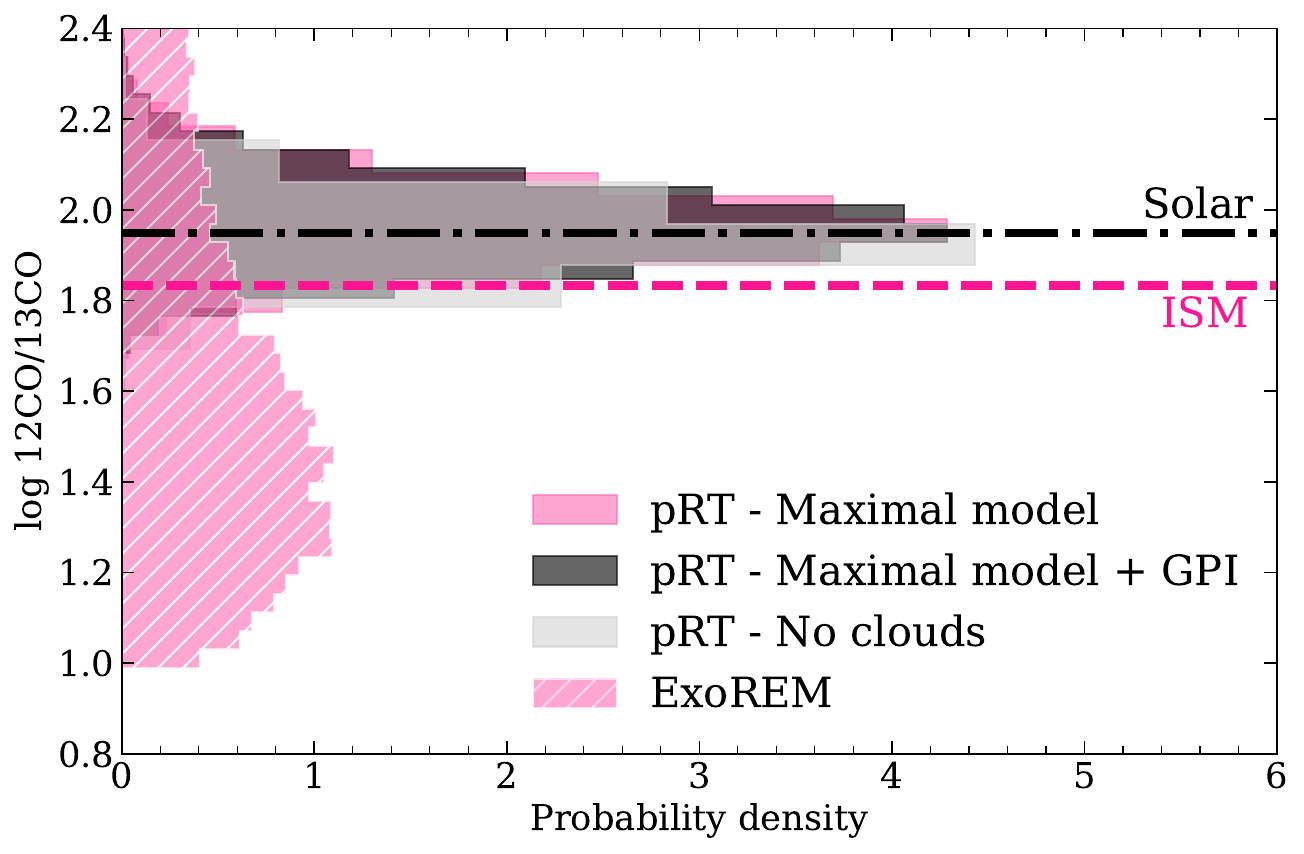}
    \caption{Posterior distributions of $\log\,^{12}\mathrm{CO}/^{13}\mathrm{CO}$ for each model, obtained with pRT (solid) and {\tt ExoREM} (hatched). The distributions are clipped at $\pm 3\sigma$. It shows agreement with ISM and solar values for all three retrieval models and a lower ratio obtained by {\tt ExoREM}, which we disregard due to a decreased fit quality of this self-consistent (less flexible) model.} 
    \label{fig:ratios}
\end{figure}

Our retrieval analysis shows a good fit for our best-fit model to the data, particularly in the $^{13}$CO absorption region (see \autoref{fig: best_fit}). When comparing our maximal model against our model without $^{13}$CO, a flux difference in the $^{13}$CO absorption region from 2.34\,µm to 2.40\,µm is apparent, with the inclusion of $^{13}$CO adding additional necessary absorption. Model comparison yields a difference in Bayesian Information Criterion (BIC) of $\Delta \text{BIC} \approx 25$, indicating strong preference for $^{13}$CO \citep{kass1995}. We retrieved a $^{12}\text{CO}/^{13}\text{CO}$ of 91$^{+24}_{-17}$ (log value of $1.96^{+0.10}_{-0.09}$, see \autoref{fig:ratios}), consistent with both the local interstellar medium (ISM) value of $68\pm15$ \citep{milam2005} and the solar value of $\sim 89$ \citep{wilson1994}. Our detection is tentatively supported by the cross-correlation function result, which returned a CCF signal of 3.6 (see \autoref{fig: templates retrieval}). We ran two additional retrievals; one without clouds and one including GPI Y-, J-, and H-band data \citep{chilcote2017}. Both runs yield an isotopologue ratio consistent with our maximal model (87$^{+20}_{-15}$ and 91$^{+24}_{-17}$, respectively). Although the no-cloud model is preferred ($\Delta \text{BIC} \approx 73$), we adopt the maximal model, as clouds are expected in $\beta$ Pic b's atmosphere, and K-band data do not constrain cloud parameters well \citep{landman2024}. The agreement with the GPI-including retrieval likely reflects the S/N dominance of GRAVITY+, however, we describe these results in \autoref{app: GPI data}.

\begin{figure}[htp]
    \centering
    \includegraphics[width=\linewidth]{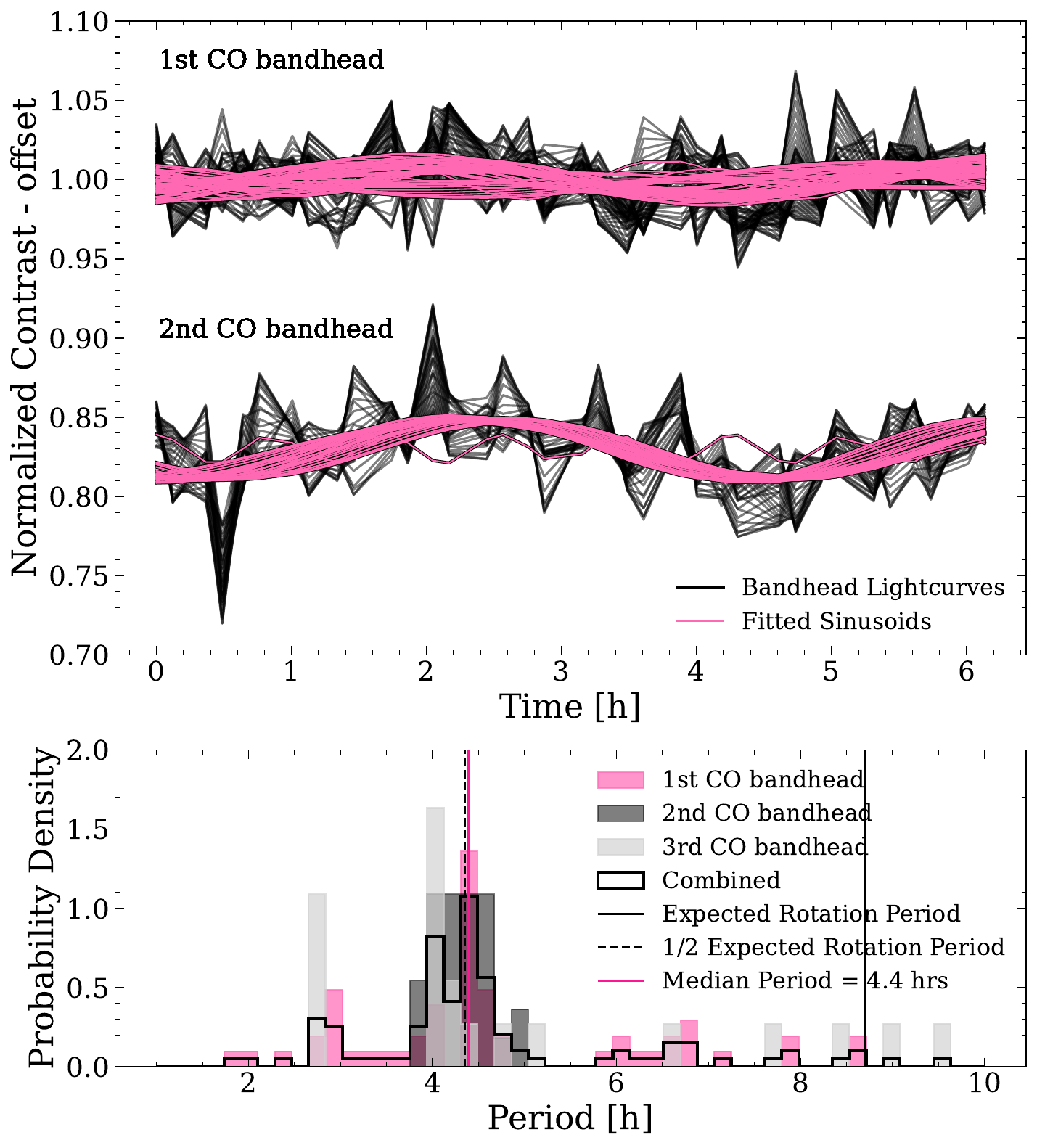}
    \caption{\textit{Top panel:} shows contrast lightcurves of 1st and 2nd CO bandhead wavelengths with kernel smoothing applied. The pink lines show the fitted sinusoids to each bandhead. The 3rd bandhead is shown in \autoref{fig:third_bandhead}.\textit{Bottom panel:} shows a histogram of the distribution of fitted periods between 0 and 10\,hrs. The probability density for each bandhead is shown, as well as the total distribution of all bandheads. The expected rotation period and its two smaller harmonics are denoted by black vertical lines. The median period is shown in pink and coincides with P/2.}
    \label{fig:variability}
\end{figure}

\subsection{Variability}
\label{sec: Variability results}

\autoref{fig:variability} shows the first two $^{12}$CO bandhead light curves with their sinusoidal fits. Fits are performed per wavelength, with periods and amplitudes derived from the median and 1$\sigma$ percentile range of the best-fit parameter distributions. The first and second bandheads show well-constrained signals, with periods and amplitudes of 4.4$^{+2.3}_{-1.4}$~hrs and 0.9$^{+0.4}_{-0.4}$\%, and 4.3$^{+0.3}_{-0.3}$~hrs and 1.92$^{+0.09}_{-0.17}$\%, respectively, which are both consistent with half the expected rotation period (P/2 $\approx$ 4.35~hrs). The third bandhead (\autoref{fig:third_bandhead}) yields a similar median period (4.5$^{+1400}_{-0.6}$~hrs) and amplitude (1.6$^{+110}_{-0.3}$\%), but with highly skewed distributions and extreme outliers, likely due to increased noise beyond 2.35\,µm from thermal background and residual tellurics. Combining all three bandheads gives 4.4$^{+2.3}_{-0.8}$~hrs and 1.4$^{+0.6}_{-0.7}$\%, though uncertainties are affected by the third bandhead. If the variability is dominated by the planet, two interpretations arise. Our 7~hr observations do not span the full $\approx 8.7$~hr rotation period, yet we find a period consistent with $P/2$. We have to treat this with caution as \cite{vanderplas2018} discusses the possibility of periodograms, and therefore sinusoidal fits, picking up low-integer harmonics, over the true signal. Alternatively, the shorter period may reflect atmospheric structure, such as features at opposing longitudes rotating in and out of view. Given the limited temporal coverage and strong telluric and instrumental corrections, confirming a planetary origin requires further analysis, which we will present in future work.

\section{Discussion and conclusion}
\label{sec:Conclusion}

We present an analysis of the atmosphere of $\beta$ Pictoris b using GRAVITY+ data. Our retrievals provide strong statistical evidence for $^{13}$CO when compared to the model excluding it, supported by a CCF signal with SNR = 3.6, indicating the absorption feature is well captured. The best fit model yields $^{12}\text{CO}/^{13}\text{CO} = 91^{+24}_{-17}$, consistent with ISM and solar values. This is in agreement with upcoming high-resolution results from \citet[companion paper]{gonzalezpicos2026} using CRIRES+. While our {\tt ExoREM} analysis does not recover this detection, this likely reflects its resolution-specific issues and the limited flexibility of self-consistent models.

Models of protoplanetary disks show that the $^{12}\text{C}/^{13}\text{C}$ ratio could vary across different reservoirs \citep{woods2009, lee2024, bergin2024}. One possible formation scenario for $\beta$~Pic~b could be accretion from a range of regions and reservoirs that average out, yielding an overall ISM- or solar-like ratio, rather than reflecting a single chemically distinct source. However, this scenario would require specific accretion conditions, balancing contributions of different disk reservoirs, and may be unlikely due to its ``fine tuning'' nature. This is further limited by the lack of constraints on the system’s natal disk and the planet’s migration history. \cite{gonzalespicos2025b} propose that the lower $^{12}\text{CO}/^{13}\text{CO}$ ratios in younger (metal-rich) M-dwarfs corroborate their later formation. This is because the ISM has become progressively enriched in $^{13}\text{C}$ over time due to galactic chemical evolution \citep{karakas2014}. While most Solar System objects (\citealt{nomura2022} and references therein) retain the higher solar ratio ($\sim89$) from 4–5 Gyr ago, $\beta$ Pic b’s young age ($\sim$23 Myr; \citealt{mamajek2014}) would suggest a value closer to the present-day ISM, such that the derived $^{12}\text{CO}/^{13}\text{CO}$ may simply reflect the bulk $^{12}\text{CO}/^{13}\text{CO}$ of its host star's natal cloud. We note, however, that its inferred $^{12}\text{CO}/^{13}\text{CO}$ is in agreement with both ISM and solar values, and the margin between these values is small. Nonetheless, we can confidently reject a $^{13}$CO enriched $^{12}\text{CO}/^{13}\text{CO}$ value. Overall, we conclude that it is still difficult to utilise $^{13}$CO as a formation tracer of giant planets, due to the uncertainties that still persist in the models and measurements.

Due to instrumental and telluric noise, extracting $\beta$ Pic b’s potential variability signal was challenging. However, after applying a novel telluric correction and a second-order polynomial to account for low-order instrumental effects, residual variability persisted in the three $^{12}$CO bandheads. Preliminarily, we derive a median period of 4.4$^{+2.3}_{-0.8}$~hrs and an amplitude of 1.4$^{+0.6}_{-0.7}$\%. The period is consistent with half the expected rotation period (P/2 $\approx 4.35$\,hrs), though uncertainties are increased by thermal and telluric noise beyond 2.35\,µm. The amplitude aligns with the upper limits for early L-type objects \cite{crossfield2014}. While suggestive of planet variability, the signal remains uncertain given the limited temporal baseline and sensitivity to residual systematics. Further observations, such as upcoming JWST work (Zhou et al in prep.), will be needed to confirm its origin.
 
\begin{acknowledgements}
This work is based on observations collected at the European Southern Observatory under GRAVITY+ GTO program ID 114.27JS (PI: L. Kreidberg). We would like to acknowledge Nicolas Pourré, who contributed significantly to this project, but has since left astronomy. Additionally, we would like to acknowledge Ewine van Dishoeck for an insightful discussion regarding planet-forming disks and isotopologues. This work benefited from the 2025 Exoplanet Summer Program in the Other Worlds Laboratory (OWL) at the University of California, Santa Cruz, a program funded by the Heising-Simons Foundation and NASA. J.W.X is thankful for support from the Heising-Simons Foundation 51 Pegasi b Fellowship (grant \#2025-5887). All reduced data and best fit models are available at \url{https://doi.org/10.5281/zenodo.20608083}
\end{acknowledgements}

\bibliographystyle{aa} 
\bibliography{betapicb} 


\begin{appendix} 

\section{Additional figures and tables}
\label{appendix}

\subsection{Telluric correction}
\label{app: telluric_correction}

The $^{13}$CO signal and the variability signal are both highly sensitive and can be significantly affected by telluric contamination in our observations. If not properly corrected, these signals may be indistinguishable from the planetary signals. The current method for estimating the stellar coherent fluxes during the planet observations involves averaging the absolute coherent fluxes of the bracketing on-star observations. However, this does not take into account the airmass trend, which will be larger towards the start and the end of the night, which, according to Beer–Lambert’s law, traces the telluric transmissivity.
To address this, we employ a novel telluric correction technique first introduced by \cite{sauter2026}, for the same observations. The correction is performed using a linear interpolation between the two adjacent stellar observations as follows:
\begin{equation}
\resizebox{\columnwidth}{!}{$
\begin{aligned}
\left| \Gamma_{\text{Star planet}} \right| &=
w \cdot \left(
\frac{
(\left| \Gamma_{\text{Star 2}} \right| - \left| \Gamma_{\text{Star 1}} \right|)
(\text{AM}_{\text{planet}}-\text{AM}_{\text{Star 2}})
}{
\text{AM}_{\text{Star 2}}-\text{AM}_{\text{Star 1}}
}
+ \left| \Gamma_{\text{Star 2}} \right|
\right) \\
&\quad + (1-w)\cdot \left| \Gamma_{\text{average}} \right|
\end{aligned}
$}
\end{equation}
where $\Gamma_{\text{Star 1}}$ and $\Gamma_{\text{Star 2}}$ are the coherent flux values of the on-star observations before and after the a planet observation. Similarly $\text{AM}_{\text{Star 1}}$ and $\text{AM}_{\text{Star 2}}$ denote the airmass values for the on-star observations and $\text{AM}_{\text{planet}}$ for the planet observations. Lastly $w$ defines a weighting factor that accounts for small changes in airmass to avoid numerical instabilities:
\begin{equation}
    w = \frac{(\text{AM}_{\text{Star 2}}-\text{AM}_{\text{Star 1}})^{2}}{\delta^{2} + (\text{AM}_{\text{Star 2}}-\text{AM}_{\text{Star 1}})^{2}}
\end{equation}
Here $\delta$ was empirically chosen to be 0.03. It has to be noted due to no star observations being taken after the last two planet observations, this data reduction only utilises 36 of the on-planet observations instead of the 38 available ones.

\subsection{Mean averaging spectra}
\label{app: mean_averaging}

As our observations span $\approx7$\,hrs with multiple epoch spectra, we combine these to boost our SNR for the retrieval analysis. Prior to averaging the spectra, outlier removal is done through sigma clipping using a 4$\sigma$ tolerance. Assuming a normal distribution of the errors, this would include 99.9937\% of the data. The mean flux in each wavelength bin is then computed through element-wise mean averaging, which is done by averaging all flux values from all spectra that fall inside that bin, not including the outliers:
\begin{equation}
\overline{S} = \frac{1}{N} \sum^N_{i=1} S_i
\end{equation}
where N is the total number of spectra, i the spectral index, and S the flux value for a specific wavelength bin. The corresponding mean covariance matrix is computed using the following equation:
\begin{equation}
\text{Cov}\left[ \overline{x} \right]=\frac{1}{N^2}\sum^N_{i=1}\text{Cov}\left[ x_i \right] =\frac{1}{N^2}\sum^N_{i=1}\Sigma_i
\end{equation}
Here it has to be noted that the $1/N$ division is applied with awareness of the outlier removal. After sigma clipping, the number of contributing flux values can differ between wavelength bins. For a given bin, the average is therefore computed using only the remaining (non-clipped) samples. This way, we were able to maintain the full wavelength range while still removing spurious data points, which results in those bins to have slightly overestimated uncertainties compared to bins that had no outliers removed.

\subsection{Retrieval forward model}
\label{app:forward model}

We define our forward model as follows:
The pressure-temperature (P-T) structure is based on \cite{zhang2025}. This is a P-T parametrisation first introduced in \cite{zhang2023}, where the atmosphere is divided into ten logarithmically spaced pressure layers, going from $10^{3}$ to $10^{-6}$, and then between these points is interpolated quadratically. This model therefore introduces eleven new free parameters. Ten of these represent the temperature gradient $d\ln{T}/d\ln{P}$ at the pressure layers, and one a reference temperature (T$_{\text{ref}}$) at a given pressure. For the seven gradients positioned at highest pressures, the priors are based on the 1-$\sigma$ confidence values of the distributions derived from the self-consistent P-T profiles of {\tt SONORA DIAMONDBACK} \citep{morley2024}. As the profiles in \cite{morley2024} do not reach upper atmosphere pressures of $10^{-4}$, the final three layers extending to $10^{-6}$ are able to vary freely, as described in \cite{molliere2025}. As clouds have been shown to have an impact on the spectrum of $\beta$ Pic b  \citep{nowak2020, ravet2025}, we kept scattering on for our retrievals and attempted to retrieve cloud properties implemented in pRT based on \cite{ackerman&marley2001}, with the clouds being parametrised the same way as in \cite{molliere2025}.

The chemical abundances were mostly interpolated using chemical equilibrium tables already included in pRT, calculated with {\tt easyCHEM} \citep{lei2024}, especially for any species we expected to be only background contributors. For these species, we retrieved the atmospheric C/O ratio and metallicity. However, for the species we expected to majorly impact the shape and features of the spectrum, we separately retrieved the mass fractions. This included CO$_2$,$^{13}$CO, $^{12}$CO, H$_2$O, and CH$_4$, which are most relevant in the K-band. The values for the atmospheric metallicity and C/O ratio were then calculated using the absorber mass fractions determined from chemical equilibrium abundances and the freely retrieved abundances, with resulting values shown in \autoref{tab: priors_posteriors}.

At a resolution of 4000, we had to factor in rotational broadening of the planets atmospheric lines for our model spectra. Here we used the code described in \cite{carvalho2023}. We assume the default value of 0.6 for limb darkening and split up our planet in 10 radial bins and 100 azimuthal bins. While work is being done to explore differential rotation in brown dwarfs and exoplanets \citep[e.g.][]{araujo2021}, we kept the differential rotation at 0, as it has not been measured for $\beta$ Pic b so far, and the inclination of the planet remains unclear.

The sources of the included opacities are the following: the collision-induced absorption opacities H$_2-$H$_2$ \citep{borysow2001, borysow2002} and H$_2-$He \citep{borysow1988, borysow1989a, borysow1989b}. The Rayleigh scattering opacities H$_2$ \citep{dalgarno1962} and He \citep{chan1965}. The line opacities H$_2$O \citep{polyansky2018}, CH$_4$ \citep{hargreaves2020}, CO$_2$ \citep{rothman2010}, $^{12}$CO \citep{rothman2010}, $^{13}$CO \citep{rothman2013}, Na \citep{allard2019}, K (line profiles by N. Allard, see \citealt{molliere2019b}), TiO (line lists by B. Plez, see \citealt{molliere2019b}), FeH \citep{bernath2020}, NH$_3$ \citep{rothman2013}, PH$_3$ \citep{sousasilva2015}, HCN \citep{harris2006}, H$_2$S \citep{rothman2013}.

In order to account for possible model inaccuracies and underestimated observational noise, we introduced uncertainty scaling to our retrieval method. In our retrievals, we add this scaling to our data set, as a free parameter. Here we followed the procedure described in \cite{line2015}, whereby the flux uncertainty $\sigma$ is scaled as following:
\begin{equation}
\sigma_{\text{scaled}} = [\sigma^2(\lambda)+10^b]^{1/2}
\end{equation}
This is done for the diagonal elements of the covariance matrices. If this scaling is not applied and underestimations are present from the data reduction process, the retrieval would become overly confident in the fit, and returning narrower parameter distribution widths than they should be. Therefore, we freely retrieve the b parameter, in order to estimate the scaled uncertainty. The priors for this uncertainty scaling was taken from \cite{line2015}.

\subsection{Self consistent modelling}

\subsubsection{Method}
\label{app: exorem-method}

To independently analyse our dataset, we also employ the self-consistent modelling grid {\tt ExoREM} with \texttt{ForMoSA}\footnote{\url{https://formosa.readthedocs.io/en/latest/}}, as was done in \cite{ravet2025}. By adding a self-consistent model to our analysis, especially because of the limited spectral range (GRAVITY+ data in K-band dominates even if GPI is added in the free retrievals), we may be able to more accurately constrain bulk parameters such as effective temperature. We can also test whether self-consistent models are flexible enough to detect trace species such as $^{13}$CO. {\tt ExoREM} is a radiative-convective equilibrium code with a grid specifically aimed to model young giant planetary mass companions \citep{baudino2015}, with the bulk parameters spanning $400K \leq \text{T}{\text{eff}} \leq 2000K$, $3.0 \leq\log{\text{g}}\leq 5.0$, $-0.5 \leq \text{[M/H]} \leq 2.0$, $0.10 \leq \text{C/O} \leq 0.80$ and a custom grid for $^{13}$CO. However, we have restricted these priors to a suitable range for our analysis and added relevant parameters, as described in \autoref{tab: priors_posteriors_exorem}. {\tt ExoREM} is also incorporates both iron and silicate clouds, as implemented in \citet{charnay2018}. This is particularly relevant as both \cite{nowak2020} and \cite{ravet2025}, found evidence of clouds in the atmosphere of $\beta$ Pic b, and silicate species are expected to condense in this temperature regime \citep{gao2020}. Clouds are treated with a simplified self-consistent parametrisation within this framework, which in this case means that it is driven by only the fastest micro-physical processes. This includes the vertical cloud distribution being computed by balancing sedimentation against vertical mixing, using an eddy diffusion coefficient K${\text{zz}}$, which also allows for disequilibrium chemistry. Due to the medium resolution of the data, we also included rotational broadening in the analysis, where we fixed v$\sin i$ to the value found in \cite{landman2024} (19.9 km/s). Similarly  to the uncertainty scaling in the retrievals, we can introduce a global scaling parameter ($s$) in {\tt ForMoSA} following \cite{ruffio2019}. For this the covariance is rescaled from $C$ to $s^2 C$. Rather than fitting for $s$ explicitly, we marginalise $s$ with respect to the likelihood.

\subsubsection{Results}
\label{app: exorem-results}

The self-consistent best-fit shows slightly more spread residuals in comparison to the retrieval best-fit, especially in the $^{13}$CO absorption region. When inspecting this region more closely, in the middle panel of \autoref{fig: exorem_results}, we can see that {\tt ExoREM} is not flexible enough to accurately fit the depths of the lines in this region. This could be affected by the available {\tt ExoREM} spectra are generated at R $\sim 10,000-8,000$ in the K band but are only Nyquist sampled at R $\sim 5,000-4,000$, which would in turn reduce the effective model resolution beyond $\sim$2.3\,$\mu$m relative to GRAVITY. We do attempt to oversample the {\tt ExoREM} spectra by interpolating them to the more finely spaced wavelength grid of the observations, which does pose risks but preserves the GRAVITY+ resolution. However, this could somewhat limit the ability to capture the line shapes of $^{12}$CO and $^{13}$CO and affect the signal in the cross-correlation function. Nonetheless we do get a constraint for $^{13}$CO, with a $^{12}\text{CO}/^{13}\text{CO}$ ratio of $28^{+22}_{11}$, as shown in \autoref{fig:ratios}, which would indicate a strongly enriched $^{13}$CO abundance. Given that the {\tt ExoREM} posterior distribution shows a long tail, the models do not accurately reproduce the CO line shapes and show no significant cross-correlation peak at 0,km,s$^{-1}$ , we do not consider the inferred $^{12}\text{CO}/^{13}\text{CO}$ ratio to be a robust constraint. Additionally, in \autoref{fig: exorem_results} it shows a small underestimation of flux at the bluest wavelengths, where H$_2$O absorbs. This may indicate a lack of flexibility of the grid modelling approach, or a too weak removal of tellurics, as less flux is removed than should be according to self-consistent physics.

\begin{table}[htbp]
\caption{Piors and mean posteriors with their 1$\sigma$ uncertainties from the retrievals with {\tt petitRADTRANS}. $\mathcal{N}$ defines Gaussian priors, while $\mathcal{U}$ signifies uniform priors.}
\label{tab: priors_posteriors}
\centering
\begin{tabularx}{0.5\textwidth}{l X X }
\hline
\noalign{\smallskip}
\hline
\noalign{\smallskip}
Parameter & Prior & Posteriors  \\ 
\noalign{\smallskip}
\hline
\noalign{\smallskip}
D$_p$ (pc)                                & $\mathcal{N}$(10,30)       & 22.3$^{+0.3}_{-0.3}$         \\
\noalign{\smallskip}
R$_p$ (R$_{\text{Jup}}$)                  & $\mathcal{N}$(0.5, 2)      & 1.37$^{+0.02}_{-0.02}$       \\
\noalign{\smallskip}
v$\sin{i}$ (km/s)                         & $\mathcal{U}$(0, 10)       & 16.1$^{+0.9}_{-0.6}$       \\ 
\noalign{\smallskip}
log g (cgs)                               & $\mathcal{U}$(2, 5.5)      & 3.59$^{+0.05}_{-0.05}$       \\
\noalign{\smallskip}
{[}M/H{]}$^*$                             & $\mathcal{U}$(-1.5, 2.5)   & -0.01$_{-0.03}^{+0.03}$       \\
\noalign{\smallskip}
C/O$^*$                                    & $\mathcal{U}$(0.1, 1.6)    & 0.65$_{-0.01}^{+0.01}$       \\
\noalign{\smallskip}
log P$_{\text{Quench}}$                    & $\mathcal{U}$(-6, 3)       & -1.9$^{+2.2}_{-2.0}$
\\
\noalign{\smallskip}
b$_{\text{GRAVITY+}}$                      & \makecell[l]{$\mathcal{U}(\log_{10}(0.01\,\sigma^2_\mathrm{min}),$\\$\log_{10}(100\,\sigma^2_\mathrm{max}))$}      & -32.30$^{+0.01}_{-0.01}$       \\ 
\noalign{\smallskip}
\hline
\noalign{\smallskip}
T$_{\text{bottom}} (K)$                   & $\mathcal{U}$(100, 8900)   & 6970$^{+180}_{-170}$ \\
\noalign{\smallskip} 
(d$\ln$T/d$\ln$P)$_1$                     & $\mathcal{N}$(0.25, 0.025) & 0.25$^{+0.01}_{-0.01}$        \\
\noalign{\smallskip}
(d$\ln$T/d$\ln$P)$_2$                     & $\mathcal{N}$(0.15, 0.03)  & 0.15$^{+0.01}_{-0.01}$        \\
\noalign{\smallskip}
(d$\ln$T/d$\ln$P)$_3$                     & $\mathcal{N}$(0.18, 0.045) & 0.18$^{+0.01}_{-0.01}$        \\
\noalign{\smallskip}
(d$\ln$T/d$\ln$P)$_4$                     & $\mathcal{N}$(0.21, 0.06)  & 0.13$^{+0.01}_{-0.01}$        \\
\noalign{\smallskip}
(d$\ln$T/d$\ln$P)$_5$                     & $\mathcal{N}$(0.16, 0.05)  & 0.04$^{+0.01}_{-0.01}$        \\
\noalign{\smallskip}
(d$\ln$T/d$\ln$P)$_6$                     & $\mathcal{N}$(0.08, 0.025) & 0.04$^{+0.01}_{-0.01}$        \\
\noalign{\smallskip}
(d$\ln$T/d$\ln$P)$_7$                     & $\mathcal{N}$(0.07, 0.02)  & 0.03$^{+0.01}_{-0.01}$        \\
\noalign{\smallskip}
(d$\ln$T/d$\ln$P)$_8$                     & $\mathcal{U}$(0.0, 0.1)    & 0.04$^{+0.03}_{-0.02}$        \\
\noalign{\smallskip}
(d$\ln$T/d$\ln$P)$_9$                     & $\mathcal{U}$(0.0, 0.1)    & 0.04$^{+0.03}_{-0.02}$        \\
\noalign{\smallskip}
(d$\ln$T/d$\ln$P)$_{10}$                  & $\mathcal{U}$(0.0, 0.1)    & 0.05$^{+0.02}_{-0.02}$        \\
\noalign{\smallskip}
\hline
\noalign{\smallskip}
$\log_{10}$X$^{\text{CO}_{2}}$            & $\mathcal{U}$(-10, -1)      & -3.97$^{+0.08}_{-0.09}$       \\
\noalign{\smallskip}
$\log_{10}$X$^{^{13}\text{CO}}$           & $\mathcal{U}$(-10, -1)      & -4.14$^{+0.09}_{-0.10}$       \\
\noalign{\smallskip}
$\log_{10}$X$^{^{12}\text{CO}}$           & $\mathcal{U}$(-10, -1)      & -2.22$^{+0.03}_{-0.03}$       \\
\noalign{\smallskip}
$\log_{10}$X$^{\text{H}_{2}\text{O}}$     & $\mathcal{U}$(-10, -1)      & -2.62$^{+0.03}_{-0.03}$       \\
\noalign{\smallskip}
$\log_{10}$X$^{\text{CH}_4}$              & $\mathcal{U}$(-10, -1)      & -7.93$^{+0.98}_{-0.99}$       \\
\noalign{\smallskip}
log K$_{\text{zz}}$                       & $\mathcal{U}$(5, 15)       & 9.4$^{+2.5}_{-2.3}$        \\
\noalign{\smallskip}
log r$^{\text{Fe}}$                       & $\mathcal{U}$(-6, 6)       & -2.3$^{+1.2}_{-1.4}$       \\
\noalign{\smallskip}
f$_{\text{sed}}^{\text{Fe}}$              & $\mathcal{U}$(0, 10)       & 5.1$^{+2.4}_{-2.4}$        \\
\noalign{\smallskip}
scale Fe                                  & $\mathcal{U}$(-6, 2)       & -3.2$^{+1.5}_{-1.4}$       \\ 
\noalign{\smallskip}
$\log \text{P}_{\text{base}}\text{Fe}$    & $\mathcal{U}$(9)       & -2.3$^{+2.5}_{-2.0}$       \\ 
\noalign{\smallskip}
log r$^{\text{Mg}_2\text{SiO}_4}$         & $\mathcal{U}$(-6, 6)       & -2.5$^{+1.3}_{-1.6}$       \\
\noalign{\smallskip}
f$_{\text{sed}}^{\text{Mg}_2\text{SiO}_4}$ & $\mathcal{U}$(0, 10)       & 5.1$^{+2.4}_{-2.5}$       \\ 
\noalign{\smallskip}
scale $\text{Mg}_2\text{SiO}_4$           & $\mathcal{U}$(-6, 2)       & -3.04$^{+1.7}_{-1.5}$       \\ 
\noalign{\smallskip}
$\log \text{P}_{\text{base}}\text{Mg}_2\text{SiO}_4$    & $\mathcal{U}$(-10, 9)       & -2.8$^{+2.2}_{-1.8}$       \\ 
\noalign{\smallskip}
\hline
\noalign{\smallskip}
\end{tabularx}
\tablefoot{$^*$ signifies the C/O and [M/H] values derived from the equilibrium chemistry used for background absorbers and then corrected using the free abundances.}
\end{table}

\begin{table}[]
\caption{Priors and mean posteriors with their 1$\sigma$ uncertainties from the self-consistent modelling with {\tt ExoREM}}
\label{tab: priors_posteriors_exorem}
\centering
\begin{tabular}{lll}
\hline
\noalign{\smallskip}
\hline
\noalign{\smallskip}
Parameter                           & Prior                     & Posterior                \\ 
\noalign{\smallskip}
\hline
\noalign{\smallskip}
T$_\text{eff}$ (K)                  & $\mathcal{U}$(1200, 2000) & 1604$^{+2}_{-2}$     \\
\noalign{\smallskip}
R$_\text{pl}$ (R$_\text{Jup}$)      & $\mathcal{U}$(0, 2)       & 1.680$^{+0.003}_{-0.003}$     \\
\noalign{\smallskip}
log(g) (dex)                        & $\mathcal{U}$(3.5, 4.5)   & 3.86$^{+0.02}_{-0.02}$     \\
\noalign{\smallskip}
{[}M/H{]}                           & $\mathcal{U}$(-0.5, 1)    & 0.06$^{+0.02}_{-0.02}$        \\
\noalign{\smallskip}
C/O                                 & $\mathcal{U}$(0.4, 0.65)  & 0.600$^{+0.001}_{-0.001}$ \\
\noalign{\smallskip}
log $^{12}$CO/$^{13}$CO             & $\mathcal{U}$(1.0, 2.6)   & 1.4$^{+0.3}_{-0.2}$     \\ 
\noalign{\smallskip}
RV (km/s)                           & $\mathcal{U}$(-50, 50)    & 37.3$^{+0.8}_{-0.8}$     \\ 
\noalign{\smallskip}\hline
\end{tabular}
\end{table}

\begin{figure}
\label{fig: pt retrieval}
    \centering
    \includegraphics[width=0.85\linewidth]{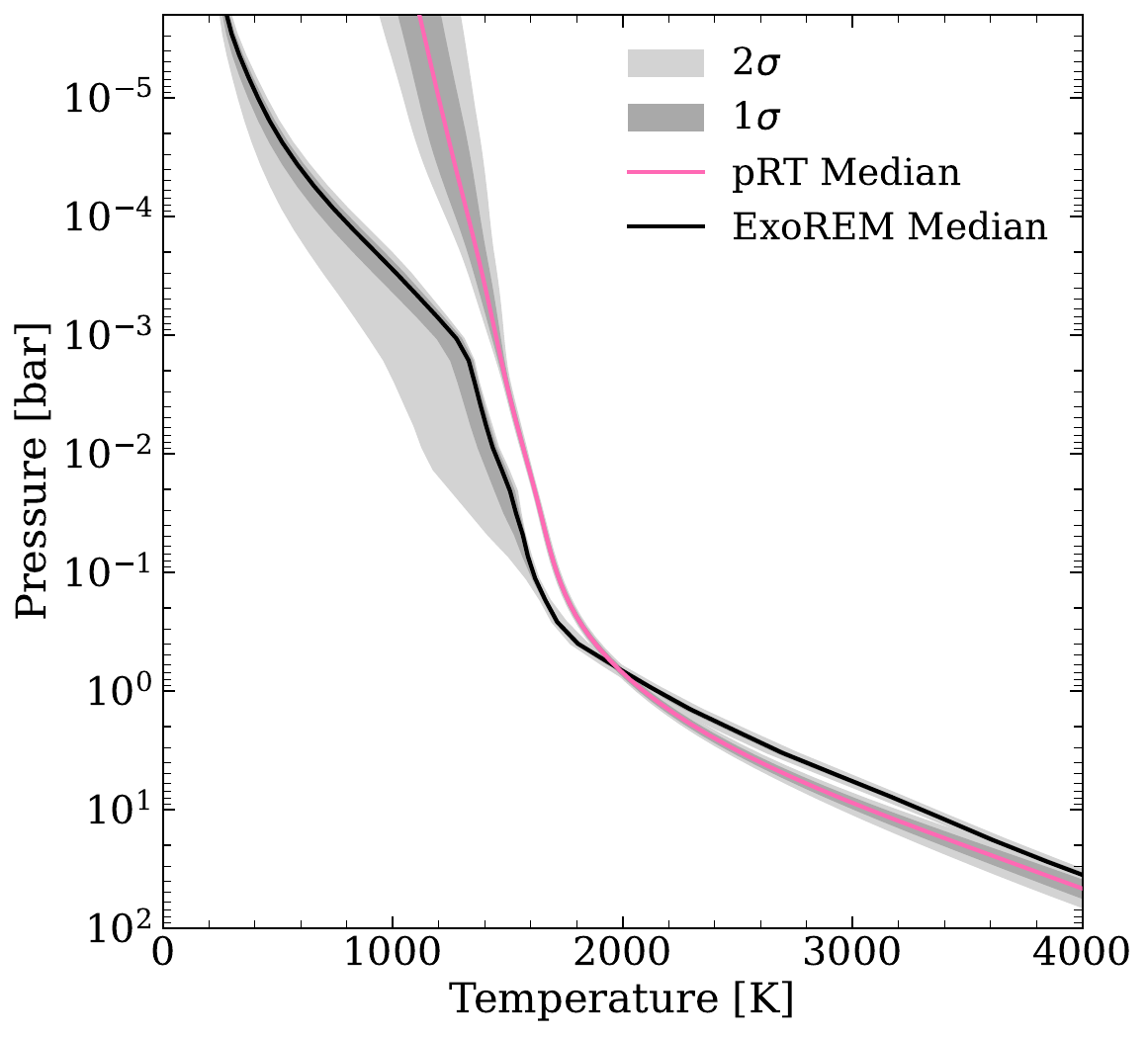}
    \caption{Displays median pressure-temperature profiles from both best fit models of {\tt petitRADTRANS} (pink) and {\tt ExoREM} (black). The grey shading shows their respective 1- and 2$\sigma$  uncertainty envelopes.}
    \label{fig:PT_prof_retrieval}
\end{figure}

\begin{figure}
    \centering
    \includegraphics[width=\linewidth]{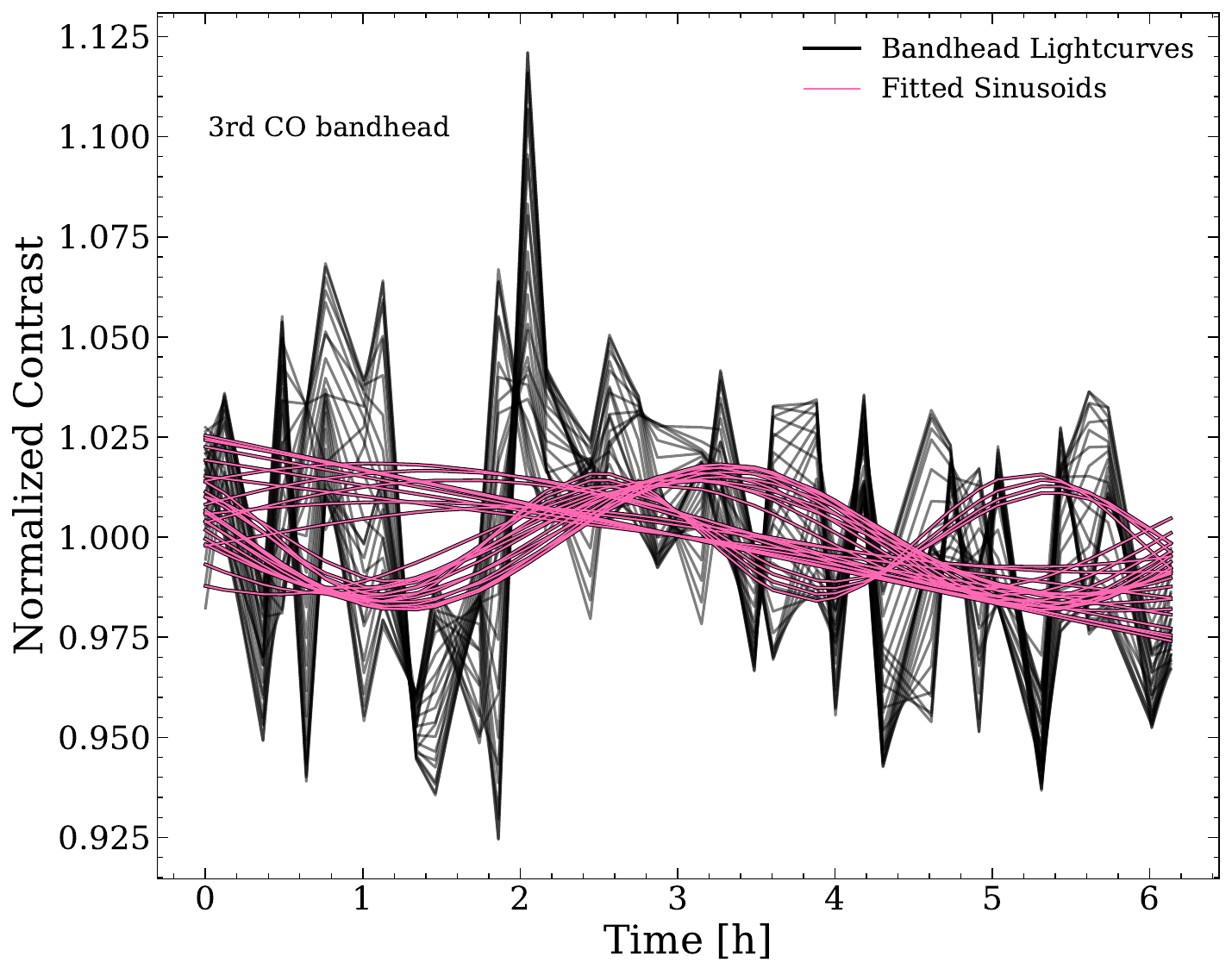}
    \caption{Contrast lightcurves of third CO bandhead wavelengths with kernel smoothing applied. The pink lines show the fitted sinusoids to the lightcurves. The 1st and 2nd Bandhead are shown in \autoref{fig:variability}.}
    \label{fig:third_bandhead}
\end{figure}

\begin{figure*}[!htp]
\label{fig: templates retrieval}
    \centering
    \includegraphics[width=\linewidth]{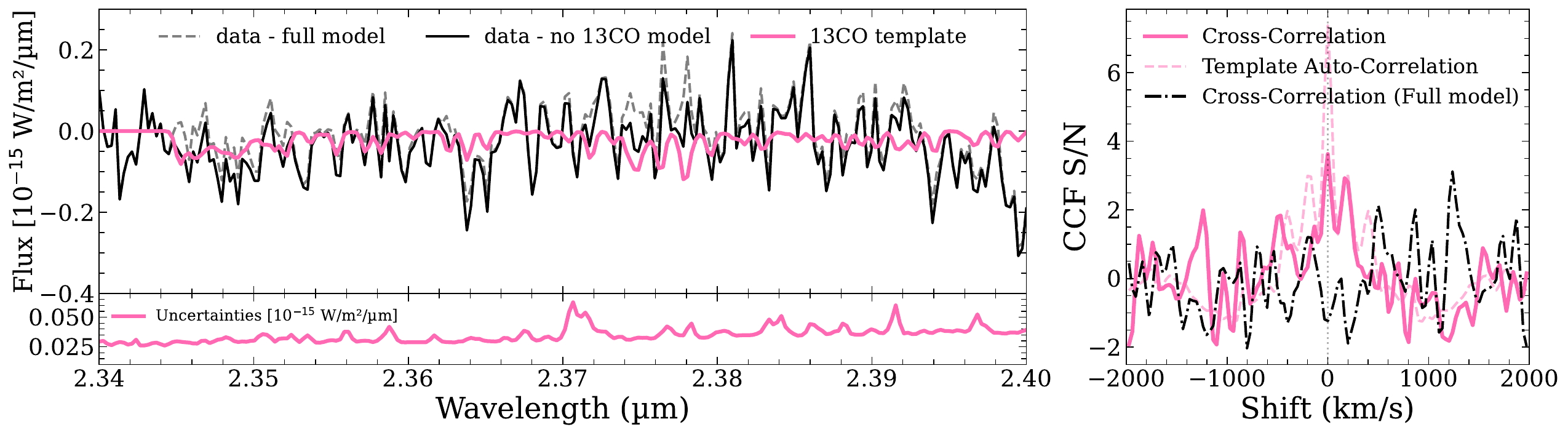}
    \caption{\textit{Top panel:} Shows the cross correlation templates created from the retrieved models. In black are the observational residuals, which are the data and the best-fit model, with $^{13}$CO turned off, subtracted. In pink we have the best fit model, subtracted with that same model but with just $^{13}$CO manually turned off, which gives the $^{13}$CO template. The grey dashed line shows the residuals when subtracting the full model from the data. \textit{Bottom panel:} Shows the observational uncertainties, by which we weighted the residuals. \textit{Right panel:} The cross-correlation function for each of the residuals with the $^{13}$CO template. The CCF between the $^{13}$CO residuals left in the data and the $^{13}$CO template is shown in pink, while black dot dashed line shows the cross-correlation between the full model residuals and the $^{13}$CO template. The autocorrelation of the $^{13}$CO signal is shown as the light pink dashed line.}
\end{figure*}

\begin{figure*}[!htp]
\label{fig: exorem_results}
    \centering
    \includegraphics[width=\linewidth]{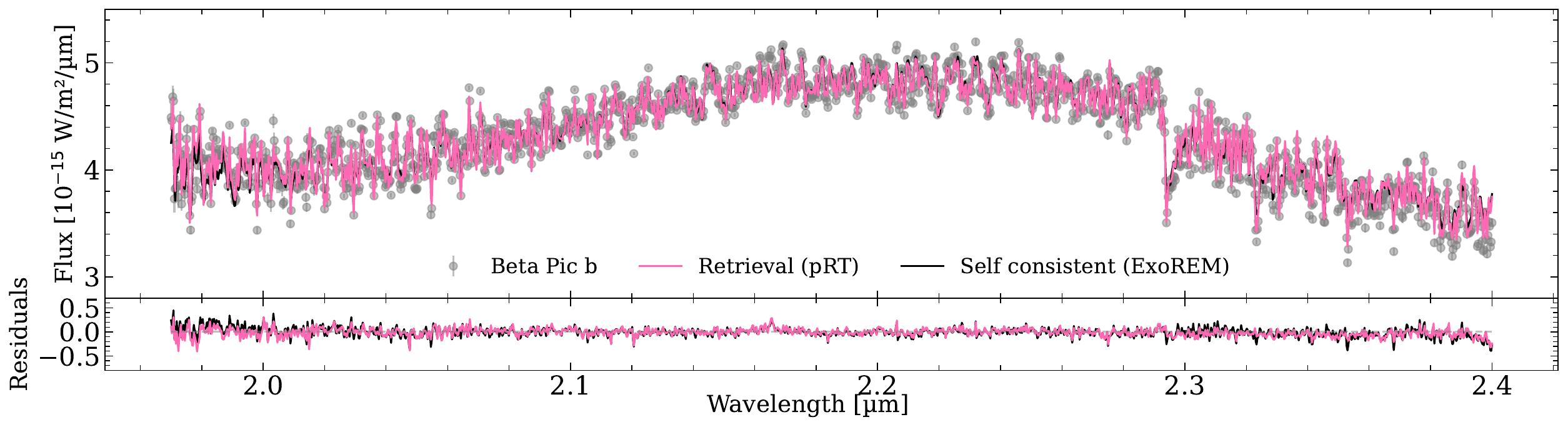}
    \includegraphics[width=\linewidth]{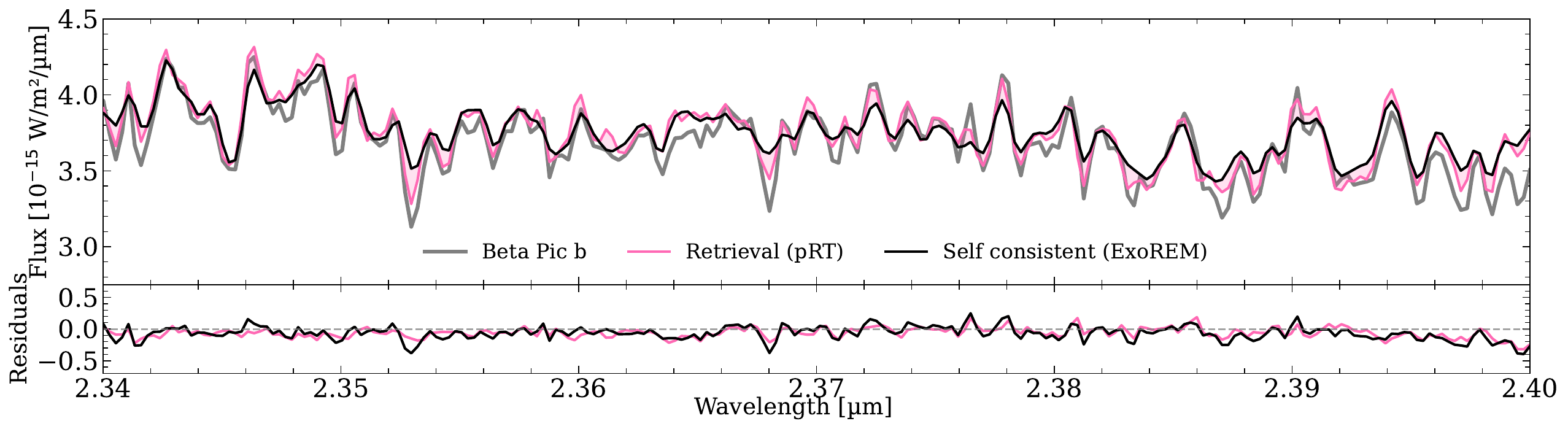}
    \includegraphics[width=\linewidth]{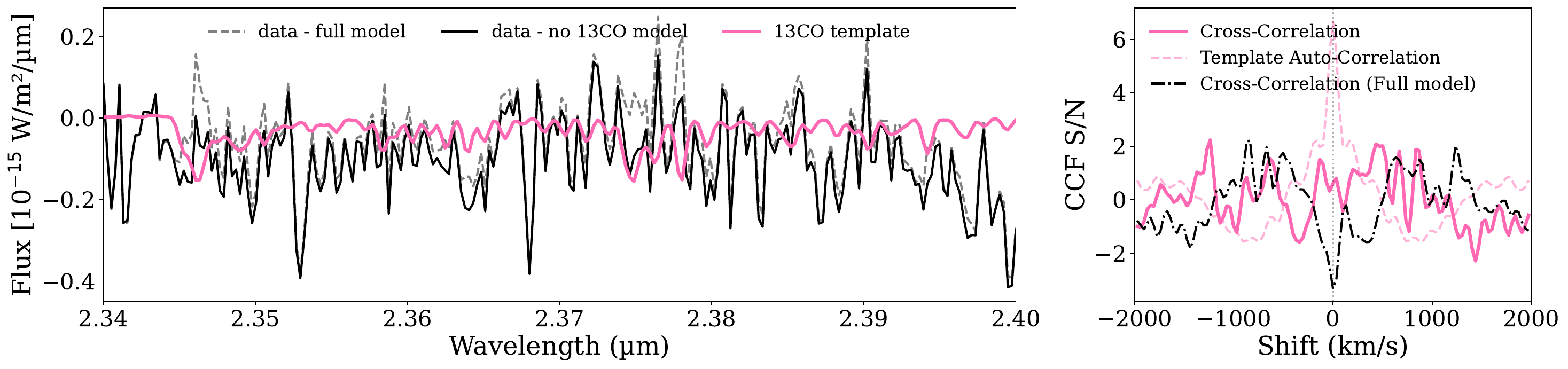}
    \caption{\textit{Top panel:} Shows the comparison between the retrieval (pink) and the self consistent model (black), as well as their respective residuals to the data. \textit{Middle panel:} Zoomed in version on the active $^{13}$CO absorption region between 2.34\,µm and 2.4\,µm. \textit{Bottom panel:} Shows the cross correlation templates for the self consistent model similarly to \autoref{fig: templates retrieval}, as well as the cross correlation function.}
\end{figure*}

\subsection{Adding GPI data}
\label{app: GPI data}

\begin{figure*}
    \centering
    \includegraphics[width=\linewidth]{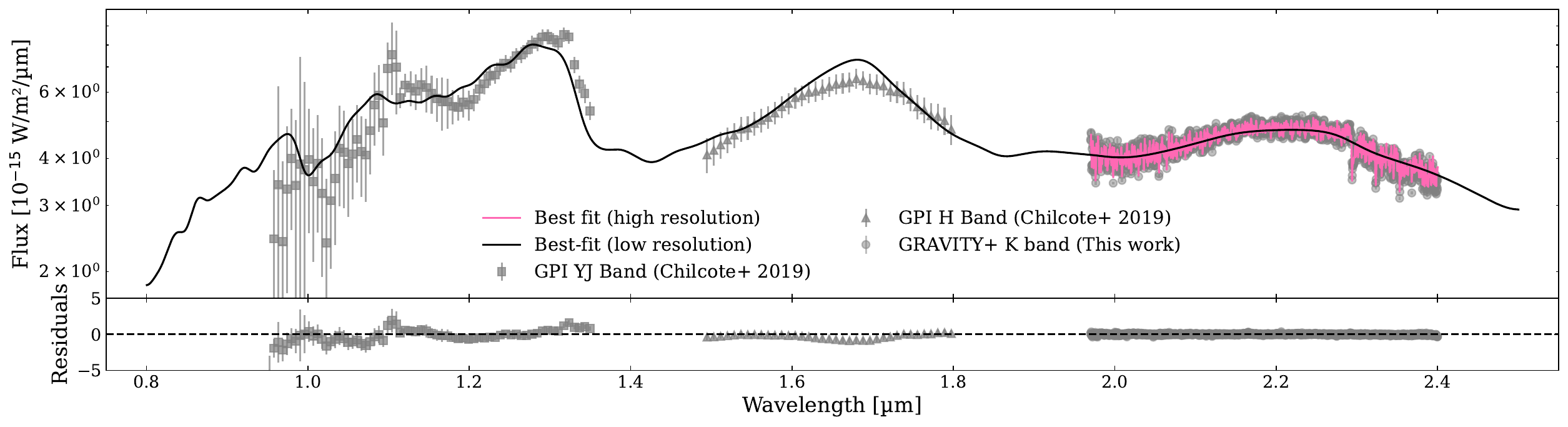}
    \caption{\textit{Top panel:} Best-fit retrieval model for the GPI and GRAVITY+ data. Grey squares and grey triangles show the GPI Y-, J-, and H-band data from \cite{chilcote2017}, respectively, and grey circles show the GRAVITY+ mean K-band data from this work, all with their respective uncertainties. The black line shows the best-fit model convolved to the GPI spectral resolution of R$\sim$70, and the pink line shows the corresponding high-resolution model. For clarity, the high-resolution model is shown only in the K-band. \textit{Bottom panel:} Shows the respective residuals between the model and data sets.}
    \label{fig:GPI_retrieval}
\end{figure*}

In our retrieval analysis, we ran our retrieval model on the Y-, J-, and H-band GPI data from \cite{chilcote2017} together with our GRAVITY+ data, in addition to the model presented in the main text. As noted by \cite{nowak2020}, while the GRAVITY K-band data are important for constraining the C/O ratio, the GPI Y-, J-, and H-band data are needed to better constrain $\log$,g. Therefore, the motivation for including the additional short-wavelength spectral information was to better constrain bulk parameters, and since there is a well-established correlation between gravity and clouds \citep{marley2012}, we hoped this might also help constrain the cloud parameters. Improvement on constraining other atmospheric parameters thereby could possibly improve the detection of $^{13}$CO or the constraint of the $^{12}$CO/$^{13}$CO ratio.
The best fit model is shown in \autoref{fig:GPI_retrieval}, where it is apparent that the quality of the fit is worse for the GPI data, which is likely caused by the GRAVITY+ data dominating the fit due to its significantly higher S/N and number of data points. This is especially shown in the H-band where the retrieval fits a much stronger triangular shape than indicated by the observed spectrum. Including these data sets, however, does appear to affect the values for $\log$,g, C/O, and metallicity (see \autoref{tab: priors_posteriors_GPI}), pushing all these values higher. This results in a significantly super-solar C/O ratio and a super-solar metallicity. On the other hand, the cloud parameters remain unconstrained, while the retrieved $^{12}$CO/$^{13}$CO ratio remains fully consistent with the GRAVITY-only results. This result might originate from the poor reproduction of the GPI spectra, meaning the retrieval cannot meaningfully exploit the additional wavelength coverage to constrain the cloud parameters, despite the GPI data nominally covering a regime where the degeneracy between $\log$,g and cloud parameters might otherwise be broken. The changes in C/O and metallicity, while statistically significant, should be interpreted with caution, if the model is unable to adequately fit the GPI data, the resulting posteriors for these parameters may reflect the retrieval compensating for systematic residuals rather than a genuine improvement in constraint. This is supported by the $^{12}$CO/$^{13}$CO ratio that appears to be almost exclusively determined by the high resolution CO features in the GRAVITY+ K-band data, which seem unaffected by the inclusion of the GPI data. We therefore consider the GRAVITY-only values of C/O and metallicity to be more reliable, and treat the GPI retrieval results primarily as a consistency check rather than an independent constraint.

\begin{table}[htbp]
\caption{Mean posteriors with their 1$\sigma$ uncertainties of bulk atmospheric parameters from the retrievals with {\tt petitRADTRANS} for both the GRAVITY+ only models and the ones including GPI data \citep{chilcote2017}, as well as the derived $^{12}$CO/$^{13}$CO ratio and effective temperature}
\label{tab: priors_posteriors_GPI}
\centering
\begin{tabular}{lll}
\hline
\noalign{\smallskip}
\hline
\noalign{\smallskip}
Parameter & GRAVITY+ & GRAVITY+GPI \\ 
\noalign{\smallskip}
\hline
\noalign{\smallskip}
R$_p$ (R$_{\text{Jup}}$)     & 1.37$^{+0.02}_{-0.02}$     & 1.37$^{+0.2}_{-0.2}$  \\
\noalign{\smallskip}
v$\sin{i}$ (km/s)            & 16.1$^{+0.9}_{-0.6}$       & 15.9$^{+0.7}_{-0.5}$  \\ 
\noalign{\smallskip}
log g (cgs)                  & 3.59$^{+0.05}_{-0.05}$     & 3.70$^{+0.06}_{-0.07}$ \\
\noalign{\smallskip}
{[}M/H{]}                    & -0.01$^{+0.03}_{-0.03}$    & 0.23$^{+0.03}_{-0.04}$   \\
\noalign{\smallskip}
C/O                          & 0.65$^{+0.01}_{-0.01}$    & 0.70$^{+0.01}_{-0.01}$   \\
\noalign{\smallskip}
$^{12}$CO/$^{13}$CO          & 91$^{+24}_{-17}$    & 91$^{+24}_{-17}$   \\
\noalign{\smallskip}
T$_{\text{eff}}$ (K)         & 1870$\pm 21$    & 1830$\pm 20$   \\
\noalign{\smallskip}
\noalign{\smallskip}
\hline
\noalign{\smallskip}
\end{tabular}
\end{table}

\begin{figure*}
    \centering
    \includegraphics[width=\linewidth]{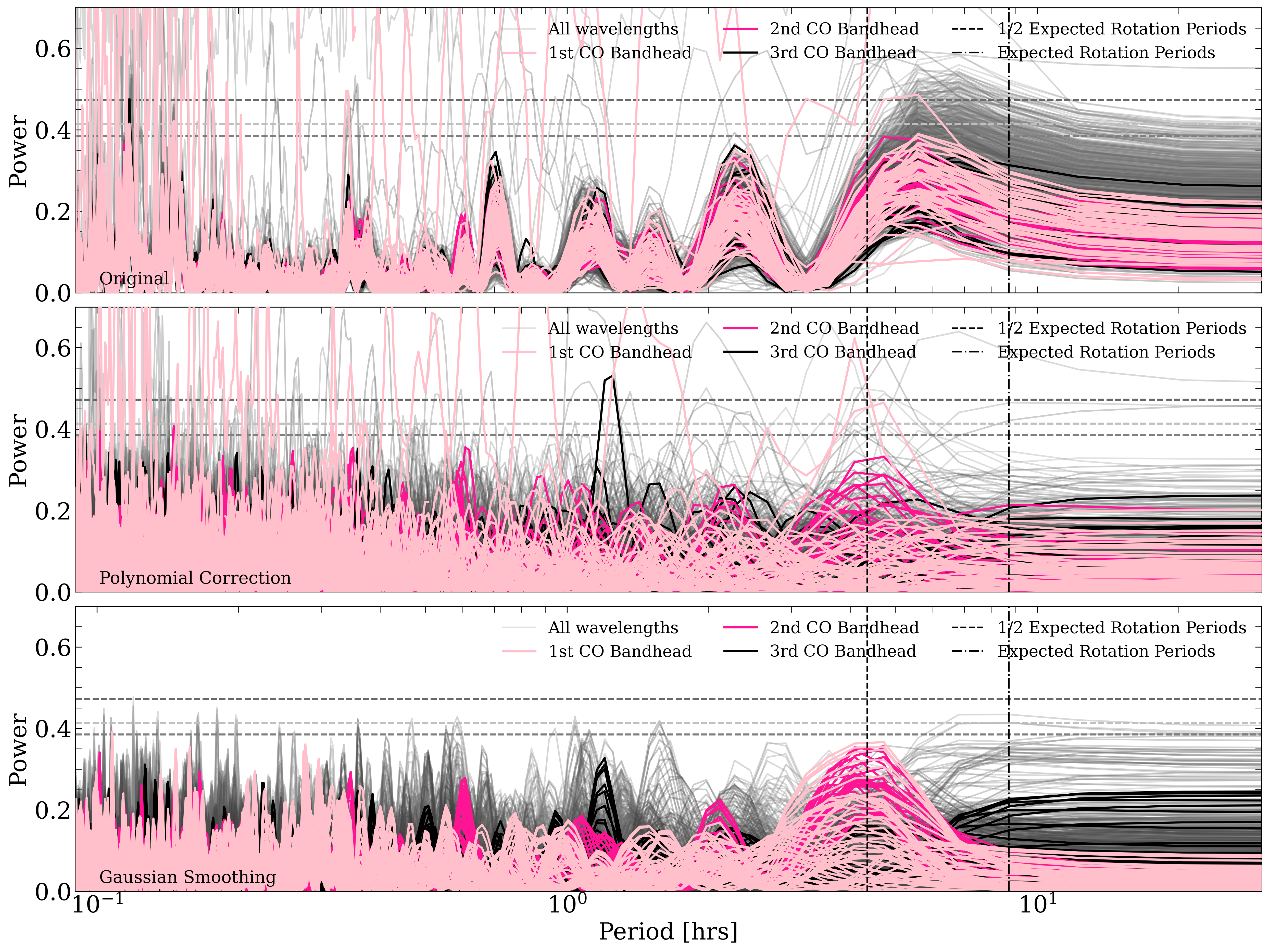}
    \caption{GLS periodograms of individual wavelength contrast lightcurves. In each of the panels, the light pink, dark pink, and black highlighted lines are the 1st, 2nd, and 3rd $^{12}$CO bandheads, respectively. The grey horizontal lines indicate the 1\%, 5\%, and 10\% false alarm probabilities. The black vertical lines indicate the expected rotation period at 8.7\,hrs and its first integer harmonic at 4.35\,hrs. The top panel shows the uncorrected periodogram, with all wavelengths significantly affected by systematic effects caused by the fibre coupling. The middle panel shows the periodogram after the polynomial correction, while the bottom panel adds Gaussian smoothing to this to highlight strong periodic signals.}
    \label{fig:lomb_scargle}
\end{figure*}
 
\end{appendix}

\end{document}